\pdfoutput=1
\documentclass[superscriptaddress,aps,twocolumn,floatfix,prb]{revtex4}
\usepackage{graphicx,amsmath,bbm}

\newcommand{\vk}{\vec k}

\newcommand{\vq}{\vec q}

\renewcommand{\vr}{\vec r}

\newcommand{\vs}{\vec s}
\newcommand{\vv}{\vec v}

\newcommand{\vsigma}{\mbox{\boldmath $\sigma$}}

\renewcommand{\vec}[1]{\mathbf{#1}}
\newcommand{\jda}{\mathop{\jmath_\alpha}\limits^{\vbox to -.5ex{\kern-0.75ex\hbox{\scriptsize$\leftrightarrow$}\vss}}}
\newcommand{\jdb}{\mathop{\jmath_\beta}\limits^{\vbox to -.5ex{\kern-0.75ex\hbox{\scriptsize$\leftrightarrow$}\vss}}}

\newcommand{\ua}{\uparrow}
\newcommand{\da}{\downarrow}

\begin{document}

\title{Spin-wave-induced correction to the conductivity of ferromagnets}
\date{\today}

\author{J.\ Danon}
\affiliation{Niels Bohr International Academy, Niels Bohr Institute, University of Copenhagen, Blegdamsvej 17, 2100 Copenhagen, Denmark}
\affiliation{Dahlem Center for Complex Quantum Systems and Fachbereich Physik, Freie Universit\"{a}t Berlin, Arnimallee 14, 14195 Berlin, Germany}
\author{A.\ Ricottone}
\affiliation{Dahlem Center for Complex Quantum Systems and Fachbereich Physik, Freie Universit\"{a}t Berlin, Arnimallee 14, 14195 Berlin, Germany}
\author{P.~W.\ Brouwer}
\affiliation{Dahlem Center for Complex Quantum Systems and Fachbereich Physik, Freie Universit\"{a}t Berlin, Arnimallee 14, 14195 Berlin, Germany}

\begin{abstract}
We calculate the correction to the conductivity of a disordered ferromagnetic metal due to spin-wave-mediated electron--electron interactions. This correction is the generalization of the Altshuler-Aronov correction to spin-wave-mediated interactions. We derive a general expression for the conductivity correction to lowest order in the spin-wave-mediated interaction and for the limit that the exchange splitting $\Delta$ is much smaller than the Fermi energy. For a ``clean'' ferromagnet with $\Delta\tau_{\rm el}/\hbar \gg 1$, with $\tau_{\rm el}$ the mean time for impurity scattering, we find a correction $\delta \sigma \propto -T^{5/2}$ at temperatures $T$ above the spin wave gap. In the opposite, ``dirty'' limit, $\Delta\tau_{\rm el}/\hbar \ll 1$, the correction is a non-monotonous function of temperature.
\end{abstract}

\maketitle

\section{Introduction}\label{sec:intr}

The electronic transport properties of disordered normal metals and their leading quantum corrections have been studied experimentally as well as theoretically for many decades, and are by now well understood.\cite{aa:book,imry:book,akkermans} Transport in {\it ferromagnetic} metals only attracted attention at a much later stage, mainly triggered by the discovery of the giant magnetoresistance,\cite{gmr1,gmr2} and the subsequent emergence of the field of spintronics.\cite{spinrev}

Although the key ingredients for understanding quantum transport in normal metals (disorder and electron--electron interactions) are also relevant for ferromagnetic metals, the magnetic order in a ferromagnet adds a significant layer of complexity and leads to additional quantum corrections to the electronic properties, that are qualitatively different from those known for normal metals. For example, spin--orbit interaction can couple the effective exchange field inside the magnet to the orbital motion of the conduction electrons, leading to a strong dependence of the conductivity on the orientation of the magnetization.\cite{PhysRevB.64.144423,kn:adam2006b} Electronic scattering off domain walls can affect the resistivity of a ferromagnet\cite{PhysRevLett.78.3773,PhysRevLett.79.5110} as well as lead to electronic dephasing.\cite{takane2} The effective interaction between conduction electrons mediated by spin waves---low-energy bosonic excitations of the magnetization---is also known to cause electronic dephasing\cite{JPSJ.72.1155,muttalibwoelfle,PhysRevB.84.224433} and changes the electronic density of states in the ferromagnet.\cite{1367-2630-15-12-123036}

Here, we address one of the most fundamental electronic properties of a metal, its d.c.\ electric conductivity. The classical (Drude) conductivity of a disordered metal can be understood in terms of electronic diffusion in the random impurity potential.\cite{ashmer,akkermans} In a normal metal, the leading quantum corrections to the conductivity at low temperatures are known to stem from weak localization and electron--electron (Coulomb) interactions.\cite{aa:book} In a ferromagnet however, the intrinsic magnetic field is expected to suppress the weak localization correction very efficiently by breaking the symmetry between time-reversed electronic paths, and in typical experiments this correction is indeed not observed.\cite{kn:wei2006,kn:neumaier2008} The interaction correction, also known as ``Altshuler-Aronov correction'', includes in a ferromagnet not only the Coulomb interaction between electrons, but also the effective interaction mediated by excitations of the magnetic order (spin waves).

Scattering between electrons and spin waves usually forms a key ingredient of semi-classical theories used to understand the temperature dependence of transport in ferromagnets.\cite{PhysRev.132.542,PhysRevB.19.384,PhysRevB.66.024433} Also recent experiments studying the quantum corrections to the conductivity of two-dimensional gadolinium sheets revealed an anomalous contribution linear in temperature, which was attributed to the spin-wave-mediated electron--electron interaction.\cite{PhysRevB.79.140408} It is thus surprising that a systematic investigation of the role of spin waves for the quantum corrections to the conductivity is still missing.

In this paper, we present a calculation of the spin-wave-mediated Altshuler-Aronov correction to the conductivity of a disordered ferromagnetic metal. Starting from a general analysis within a standard perturbative approach, solely restricted by the constraint that the effective exchange splitting $\Delta$ be smaller than the Fermi energy, we present a detailed analysis of two limiting cases: (i) The ``clean limit'' $\Delta\tau_{\rm el}/\hbar \gg 1$, with $\tau_{\rm el}$ the elastic electronic scattering time, and (ii) the ``dirty limit'' $\Delta\tau_{\rm el}/\hbar \ll 1$. In the clean limit, which is the most relevant limit for elemental ferromagnets because of the largeness of $\Delta$ in that case, we find a temperature dependence of $\propto T^{d/2+1}$ for temperatures above the spin-wave gap, $d$ being the effective dimensionality of the sample. For the dirty limit we present numerical evaluations of the correction, which exhibit a non-monotonous behavior that qualitatively resembles the spin-wave-induced correction to the density of states.\cite{1367-2630-15-12-123036}

The remainder of this paper is organized as follows. In Section \ref{sec:model} we present the model we use to describe the conduction electrons, the $d$-band spin waves, and their interaction. Our calculation of the Altshuler-Aronov correction using diagrammatic perturbation theory is then described in Section \ref{sec:dos}, where we also present our most general result. Then, in Section \ref{sec:res}, we consider the limits of large and small $\Delta\tau_{\rm el}/\hbar$ separately: For large $\Delta\tau_{\rm el}/\hbar$ we arrive at analytic results, whereas for small $\Delta\tau_{\rm el}/\hbar$ we present numerical results. Finally, in Section \ref{sec:conc} we summarize and place our work in a broader context.

\section{Model}\label{sec:model}

We will employ the same model as we used in earlier work, where we calculated the spin-wave-induced electronic dephasing\cite{PhysRevB.84.224433} and the corrections to the density of states.\cite{1367-2630-15-12-123036} However, to make the present paper self-contained, we will briefly summarize the key elements of our model, and explain our microscopic description of the conduction electrons in the disordered ferromagnet and their interaction with fluctuations of the magnetization of the $d$-band electrons.

The $s$-band conduction electrons, subject to the impurity potential $V(\bf r)$, are described by the Hamiltonian
\begin{equation}
H = \epsilon_k + V(\vr) + H_{sd},
\label{eq:ham1}
\end{equation}
where $\epsilon_k =\hbar^2k^2/2m - \varepsilon_{\rm F}$ describes the kinetic energy of the electrons, measured relative to the Fermi energy $\varepsilon_{\rm F}$, and 
\begin{align}
H_{sd} = - J \vs(\vr) \cdot \vsigma,
\end{align}
describes the exchange interaction between the conduction electrons and the $d$-band electrons. Here, $J$ is the exchange constant and $\vs(\vr)$ is the spin density of the $d$-band electrons expressed in units of $\hbar$.

We rewrite the exchange term as
\begin{equation}
H_{sd} = 
- \frac{\Delta}{2}\sigma_z
-J \left( \begin{array}{cc} 0 & s_-({\bf r},t) \\ s_+({\bf r},t) & 0 \end{array} \right),
\end{equation}
where we have split the term in a part describing the effective exchange splitting $\Delta = 2 J \bar s$ resulting from the mean $d$-band magnetization $\bar s$ (which we chose to point along the $z$-axis)\cite{cond_corr_fm:1} and a part describing the coupling to fluctuations $s_{x,y}(\vr,t)$ around this mean value, where we use the notation $s_{\pm} = s_x \pm i s_y$. We then write $H_{sd}$ in a second-quantized form,
\begin{align}
  H_{sd} =  & -\frac{\Delta}{2} \sum_\vk \left[c^\dagger_{\vk,\ua}c_{\vk,\ua} - c^\dagger_{\vk,\da}c_{\vk,\da} \right]
  \nonumber\\
  & -\frac{J}{{\cal V}} \sum_{\vk,\vq} \left[c^{\dagger}_{\vk+\vq,\ua} c_{\vk,\da}
  s_{\vq,-} + c^{\dagger}_{\vk+\vq,\da} c_{\vk,\ua} s_{\vq,+} \right],
\end{align}
with $s_{\vq,\pm} = \int d\vr s_{\pm}(\vr , t) e^{-i \vq \cdot \vr}$ the Fourier transform of the spin density and ${\cal V}$ the volume of the ferromagnet.

The spin waves are characterized by the transverse spin susceptibility
\begin{equation}
 \chi^{R}_{-+}(\vq,\tau) = -\frac{1}{\hbar{\cal V}}i\Theta(\tau) \langle [s_{-\vq,-}(\tau),s_{\vq,+}(0)] \rangle,
\label{eq:spincorr}
\end{equation}
which describes the response of the $d$-electron spin density to an applied magnetic field. In this expression $\Theta(\tau)$ is the Heaviside step function. The Fourier-transformed susceptibility\cite{kubospinsus,JPSJ.72.1155}
\begin{eqnarray}
  \chi^{\rm R}_{-+}(\vq,\Omega) &=&
  \int d\tau \, \chi^{\rm R}_{-+}(\vq,\tau) e^{i \Omega \tau} \\ &=&
  \frac{ - 2 \bar{s}}{\hbar\Omega + \hbar\omega_{\vq}^{\rm sw} + i\eta},
  \label{eq:chisw}
\end{eqnarray}
is directly related to the spin-wave dispersion $\hbar\omega_{\vq}^{\rm sw}$. (Here $\eta$ is a positive infinitesimal and we assumed that $\omega_{\vq}^{\rm sw} = \omega_{-\vq}^{\rm sw}$.) The susceptibility for opposite spin orientations reads
\begin{equation}
  \chi^{\rm R}_{+-}(\vq,\Omega) =
  \frac{2 \bar{s}}{\hbar\Omega - \hbar\omega_{\vq}^{\rm sw} + i\eta}.
\end{equation}
Following Refs.\ \onlinecite{PhysRevB.79.140408,muttalibwoelfle} we take the simple phenomenological form
\begin{equation}
\hbar\omega_\vq^{\rm sw} = \hbar D^{\rm sw} q^2 + C
\label{eq:disp}
\end{equation}
for the spin-wave dispersion, where $D^\text{sw}$ is the spin wave stiffness, usually of the order $\hbar D^\text{sw} \sim \Delta/ k_F^2$, and $C$ is the spin wave gap. This gap can be due to e.g.\ an externally applied magnetic field in the $z$-direction (in which case $C = g\mu_B B$, where $B = B_{\rm ext} - g\mu_B \mu_0 \bar{s} \xi/4 \pi$ is the external magnetic field corrected for the demagnetizing field of the device), or to the magnetocrystalline anisotropy of the material in question (in which case $C = 2K/\bar{s}$, where $K$ is the energy density characterizing the anisotropy).

\section{Diagrammatic calculation}\label{sec:dos}

In this Section we will present our diagrammatic calculation of the Altshuler-Aronov correction to the conductivity resulting from spin-wave-mediated electron--electron interactions. Following our previous work,\cite{PhysRevB.84.224433,1367-2630-15-12-123036} we assume the exchange splitting always to be small in comparison to the Fermi energy, $\Delta \ll \varepsilon_{\rm F}$. This condition is essential for the validity of the impurity perturbation theory: If it were not met, then in the clean limit the interaction correction, which is expected to be small in the parameter $\hbar/\Delta\tau_{\rm el}$, would be of the same order of magnitude as $\hbar/\varepsilon_{\rm F}\tau_{\rm el}$, which is the small parameter of the diagrammatic perturbation theory. 

There are other technical reasons why the condition $\Delta \ll \varepsilon_{\rm F}$ is desired: Taking $\Delta \ll \varepsilon_{\rm F}$ implies that majority and minority electrons have the same density of states $\nu$ at the Fermi level, that they have the same Fermi velocity $v_{\rm F}$, and that they have the same mean free path for impurity scattering. (In the present calculation, we will find that for the clean case, corresponding to large $\Delta$, the corrections do not depend on $\tau_{\rm el}$, rendering it thus unnecessary to keep track of two different scattering times.)

We now describe our calculation of the lowest-order interaction corrections to the conductivity. The diagrammatic approach follows similar calculations in the literature.\cite{aa:book,doi:10.1088/0959-7174/9/2/308,PhysRevB.64.214204} The conductivity tensor
\begin{equation}
\sigma_{\alpha\beta}(\vr,\vr';\omega) = \frac{i}{\hbar\omega} \Pi_{\alpha\beta}^R(\vr,\vr';\omega) - \frac{e^2n(\vr)}{im\omega}\delta(\vr-\vr')\delta_{\alpha\beta}
\end{equation}
is expressed in terms of the Fourier transform of the current-current correlation function
\begin{equation}
\Pi_{\alpha\beta}^R(\vr,\vr';t) = -i\Theta(t) \langle [j_\alpha (\vr;t),j_\beta (\vr';0)]\rangle,
\end{equation}
and the local electronic density $n(\vr)$. The current operator $j_\alpha(\vr;t)$ is expressed in terms of electron creation and annihilation operators
\begin{align}
j_\alpha(\vr;t) & = \sum_{\sigma = \uparrow,\downarrow}
  \psi_{\sigma} (\vr;t)\jda\psi^\dagger_{\sigma} (\vr;t) \nonumber\\
& \equiv \frac{\hbar e}{2im} 
\sum_{\sigma}
\left( [\partial_\alpha \psi_{\sigma}(\vr;t) ] \psi^\dagger_{\sigma}(\vr;t)
 \right. \nonumber \\ & \left. \hspace{5em}
  \mbox{}  - \psi_{\sigma} (\vr;t) [ \partial_\alpha \psi^\dagger_{\sigma}(\vr;t) ]\right).
\end{align}
We will focus here on the spatially-averaged symmetrized conductivity of the sample, 
\begin{equation}
  \overline{\sigma}_{\alpha\beta} \equiv \frac{1}{2{\cal V}}{\rm Re}\! \int \! d\vr d\vr' \left\{\sigma_{\alpha\beta}(\vr,\vr';0) + \sigma_{\beta\alpha}(\vr,\vr';0)\right\},
\end{equation}
where ${\cal V}$ denotes the volume of the sample.

\begin{figure}[b]
\begin{center}
\includegraphics{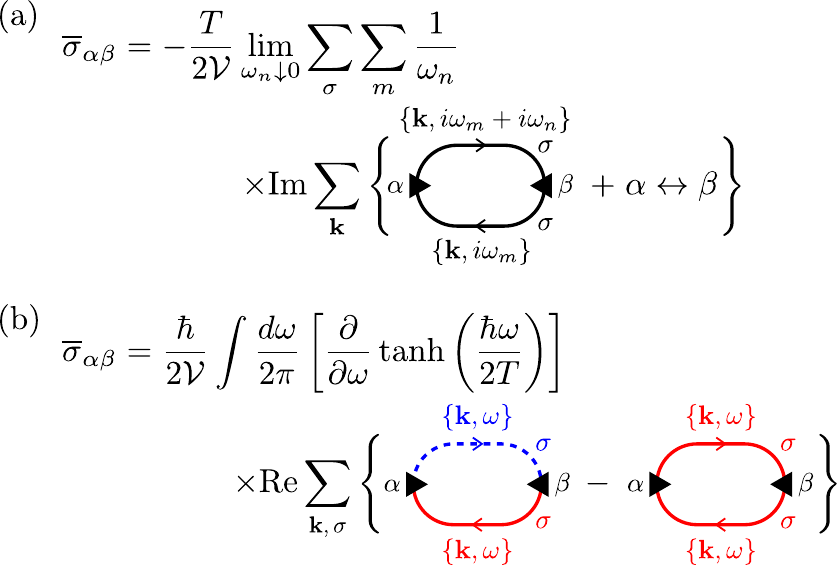}
\caption{(Color online) Diagrammatic representation of the conductivity for non-interacting electrons. (a) The conductivity in terms of temperature Green functions, of which the momentum, imaginary Matsubara frequency, and spin is indicated. The solid triangles represent the current vertices $\jda$ and $\jdb$ and give rise to factors $e\hbar k_{\alpha,\beta}/m$. (b) The same expression after analytic continuation. The blue dashed (red solid) lines represent advanced (retarded) Green functions, of which the momentum, frequency, and spin is again indicated.}\label{fig:dia1}
\end{center}
\end{figure}
Without the interaction $H_{sd}$, the conductivity can then be expressed in terms of the Matsubara Green functions ${\cal G}_\vk(i\omega_m)$,
\begin{align}
  \overline{\sigma}_{\alpha\beta} = &-\frac{\hbar^2 e^2 T}{2m^2 {\cal V}}
   \label{eq:condni}
   \lim_{\omega_n \downarrow 0} \sum_{\sigma} \sum_{m} \frac{1}{\omega_n} \\
& \times  {\rm Im}\sum_\vk  \Big\{ k_\alpha {\cal G}_{\vk,\sigma}(i\omega_m+i\omega_n)k_\beta {\cal G}_{\vk,\sigma}(i\omega_m) \nonumber\\
& \hspace{4.5em}+ k_\alpha {\cal G}_{\vk,\sigma}(i\omega_m)k_\beta {\cal G}_{\vk,\sigma}(i\omega_m+i\omega_n) \Big\}, \nonumber
\end{align}
where the sum runs over all fermionic Matsubara frequencies $\omega_m = (2m+1)\pi T/\hbar$. Here ${\cal G}_{\vk,\sigma}(i\omega_m)$ is the electronic temperature Green function with momentum $\vk$ and Matsubara frequency $i\omega_m$. In Fig.\ \ref{fig:dia1}a we show a diagrammatic representation of (\ref{eq:condni}). After analytic continuation this expression yields the standard expression for the conductivity of non-interacting electrons in terms of advanced and retarded Green functions,\cite{akkermans} as pictorially represented in Fig.\ \ref{fig:dia1}b. Averaging over disorder produces the Drude conductivity $\sigma_0$ as well as the weak localization correction to the conductivity.

\begin{figure}[t]
\begin{center}
\includegraphics{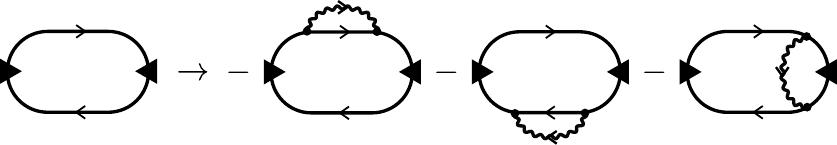}
\caption{The Fock corrections to the diagram of Fig.\ \ref{fig:dia1}a. The wiggly lines represent the interaction propagator $\chi_{-+}$.}\label{fig:dia2}
\end{center}
\end{figure}

To lowest nontrivial order, spin-wave-mediated interactions between electrons give rise to Hartree and Fock corrections to the diagrams shown in Fig.\ \ref{fig:dia1}. Due to the spin-flip nature of the spin-wave-mediated interaction we can dismiss the Hartree corrections and focus on the Fock terms. The resulting corrections to the diagrams are twofold: (i) One has to include a self-energy into the Green functions in (\ref{eq:condni}),
\begin{align}
{\cal G}_{\vk,\sigma}(i\omega_m) \to -\frac{J^2T}{{\cal V}}\sum_{\vq,l} {\cal G}_{\vk,\sigma}(i\omega_m)^2 \chi_{\sigma,-\sigma}(\vq,i\Omega_l)\nonumber\\
\times {\cal G}_{\vk-\vq,-\sigma}(i\omega_m - i\Omega_l),
\end{align}
where $\Omega_l = 2\pi l T/\hbar$ and we now included the index $\sigma$ indicating the spin of the electron. (ii) The right current vertex (labeled $\beta$) is renormalized according to
\begin{align}
&{} {\cal G}_{\vk,\sigma}(i \omega_m+i\omega_n)k_\beta {\cal G}_{\vk,\sigma}(i\omega_m) \to \nonumber\\
&\hspace{.5em}-\frac{J^2T}{{\cal V}}\sum_{\vq,l} {\cal G}_{\vk,\sigma}(i\omega_m+i\omega_n)k_\beta {\cal G}_{\vk,\sigma}(i\omega_m) \chi_{\sigma,-\sigma}(\vq,i\Omega_l) \nonumber\\
&\hspace{5.5em}\times {\cal G}_{\vk-\vq,-\sigma}(i\omega_m+i\omega_n - i\Omega_l) \nonumber\\
&\hspace{5.5em}\times {\cal G}_{\vk-\vq,-\sigma}(i\omega_m - i\Omega_l).
\end{align}
The diagram shown explicitly in Fig.\ \ref{fig:dia1}a thus acquires in total three corrections, of the type pictured in Fig.\ \ref{fig:dia2}.

After analytic continuation of the resulting expression, which is required for taking the limit $\omega_n \to 0$, one may separate the interaction correction in two sets of terms:\cite{doi:10.1088/0959-7174/9/2/308} (i) There is a ``dephasing'' correction $\delta\sigma_{\alpha\beta}^{\rm deph}$ due to the interaction-induced dephasing of coherently propagating electronic amplitudes. This dephasing involves inelastic collisions which change the energies of the conduction electrons, and it affects the weak localization correction present in the non-interacting picture. (ii) There is an ``interaction'' correction $\delta\sigma_{\alpha\beta}^{\rm int}$ which represents the elastic scattering of electrons from the self-consistent inhomogeneous potential set up by all other conduction electrons. This correction can be seen as an interaction-induced modification of the effective impurity potential felt by the electrons, thereby affecting the Drude conductivity of the non-interacting picture. The latter, the so-called Altshuler-Aronov correction, is believed to be responsible for the anomalous temperature dependence of the conductivity observed in experiment\cite{PhysRevB.79.140408} and is the one we want to calculate in the present work.
\begin{figure}[t]
\begin{center}
\includegraphics{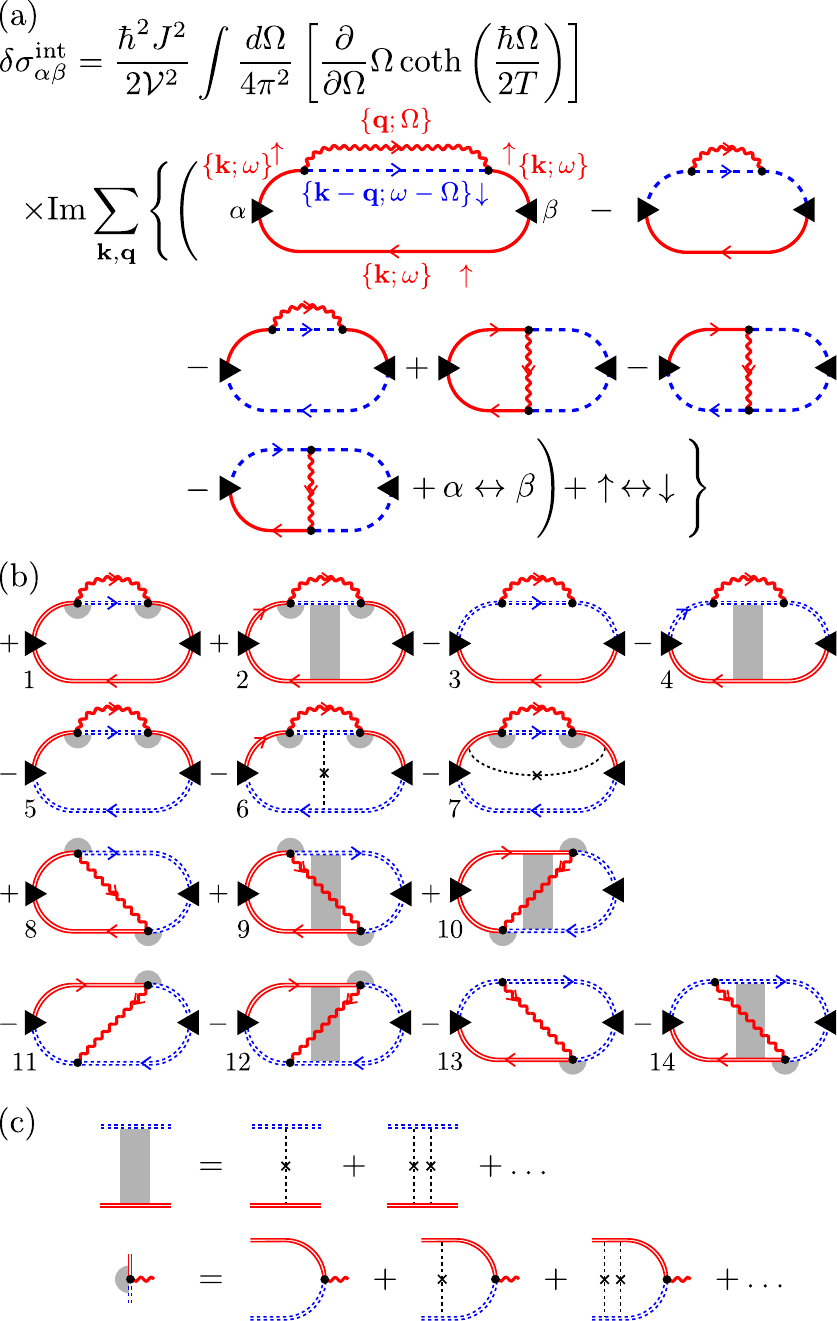}
\caption{(Color online) (a) The Fock diagrams for the interaction correction to the conductivity, after analytic continuation. Red wiggly lines represent the retarded interaction propagator $\chi^R_{-+}(\vq,\Omega)$. The exchange of spin indices $\ua\,\leftrightarrow\,\da$ indicated in the figure also includes a change of the interaction propagator $\chi_{-+} \to \chi_{+-}$. (b) The diagrams resulting from averaging the interaction correction shown in (a) over random impurity configurations. Double lines represent impurity averaged electronic Green functions and correlated scattering events are pictured by black dashed lines marked with a cross. The definition of the diffuson ladders and a renormalized current vertex (gray areas) is given in panel (c). Note that the vertex correction includes a term without impurity scattering.}\label{fig:dia3}
\end{center}
\end{figure}

In Fig.\ \ref{fig:dia3}a we show the expression for $\delta\sigma_{\alpha\beta}^{\rm int}$ after analytic continuation. We left out all diagrams containing only retarded or only advanced electronic Green functions, since these give no contribution to $\delta \sigma_{\alpha\beta}^{\rm int}$ and we used that a product of Green functions, as shown in the first diagram, does not depend on the common frequency $\omega$ as long as all relevant energies $\hbar\omega$ are of the order of the Fermi energy.

The next step is to average $\delta\sigma_{\alpha\beta}^{\rm int}$ over the disorder potential. First of all, the impurity averaged electronic Green functions acquire a finite (imaginary) self-energy,
\begin{align}
  \langle G^R_{\vk,\sigma} (\omega) \rangle & =
  \frac{1}{\hbar\omega - \epsilon_{k} + \Delta \sigma/2 + \hbar i/2 \tau_{\rm el}}, \label{eq:grav}\\
  \langle G^A_{\vk,\sigma} (\omega) \rangle & =
  \frac{1}{\hbar\omega - \epsilon_{k} + \Delta \sigma/2 - \hbar i/2 \tau_{\rm el}}. \label{eq:gaav}
\end{align}
Second, we have to include the possibility of correlated scattering events in different electronic propagators.
In Fig.\ \ref{fig:dia3}b we show all impurity averaged diagrams we obtain from the diagrams of Fig.\ \ref{fig:dia3}a. The double solid red and dashed blue lines represent the impurity averaged retarded and advanced Green functions given in (\ref{eq:grav}) and (\ref{eq:gaav}), respectively, and the black dashed lines marked with a cross indicate correlated scattering events for different Green functions, such as they appear in the Hikami box.\cite{akkermans} The diffuson ladder of correlated scattering events connecting a retarded and advanced Green function (describing classical electronic diffusion) is represented by the gray areas, see Fig.\ \ref{fig:dia3}c.

The usual approach for the calculation of the interaction correction to the conductivity is to employ the diffusion approximation: One assumes all deviations from the common energy and momentum $\hbar\omega$ and $\hbar\vk$ to be small compared to $\hbar/\tau_{\rm el}$. More precisely, for the diagrams of Fig.\ \ref{fig:dia3} one would require that $(\Omega\pm\Delta/\hbar)\tau_{\rm el} \ll 1$ (since each interaction vertex involves an electronic spin-flip) and $qv_{\rm F}\tau_{\rm el} \ll 1$. This allows to (i) neglect $\Omega\pm\Delta/\hbar$ and $\vq$ in all short-range combinations of Green functions and (ii) expand the structure factor of the diffuson in small $(\Omega\pm\Delta/\hbar)\tau_{\rm el}$ and $qv_{\rm F}\tau_{\rm el}$, leading to the standard diffusion poles.\cite{akkermans} With this approximation, one could focus exclusively on the diagrams with the largest number of diffuson ladders, and diagrams 3, 4, and 11--14 of Fig.\ \ref{fig:dia3}b could be dismissed.\cite{note_diagrams}

However, for the case of spin-wave-mediated interactions in a ferromagnet these assumptions cannot be justified in all regimes.\cite{PhysRevB.84.224433,1367-2630-15-12-123036} First, the typically large mismatch between the spin wave stiffness $D^{\rm sw}$ and the electronic diffusion constant $D \sim v_{\rm F}^2\tau_{\rm el}$ causes spin waves with small energy $\hbar\Omega$ to carry a large momentum from an electronic point of view, i.e., $(D/D^{\rm sw})\Omega\tau_{\rm el} \sim 1$ for relatively small $\Omega$. Usually, relevant interaction energies are of the order of the temperature $\hbar\Omega \sim T$, and this implies that a strictly diffusive approach would be only valid when $(D/D^{\rm sw})T\tau_{\rm el}/\hbar \ll 1$, typically restricting its validity to unrealistically low temperatures. Second, for the clean limit---which is the relevant limit for most elemental ferromagnets---, one has $\Delta \tau_{\rm el}/\hbar \gg 1$, which immediately rules out a diffusion approximation. In this case, all diffuson ladders appearing in the diagrams of Fig.\ \ref{fig:dia3}a connect two propagators with an energy difference of $\sim\Delta$. Such a pair of one advanced and one retarded electronic propagator dephases on a length scale much smaller than the electronic elastic mean free path, implying that the largest contributions come from diagrams with the {\it smallest} number of impurity lines connecting propagators of opposite spin. For the interaction correction to the conductivity, this would mean that diagrams 1, 3, 5, 7, 8, 11, and 13 of Fig.\ \ref{fig:dia3}b can be expected to provide a large contribution. (The ladders at the vertices also contain a term without impurity scattering and should therefore be kept, and diagram number 7 should be kept since its single impurity line connects propagators of the same spin species.) This clearly can lead to results very different from those obtained within the diffusion approximation. The best approach is thus to keep all diagrams of Fig.\ \ref{fig:dia3}b and retain the energy and momentum differences $\hbar\Omega\pm\Delta$ and $\hbar\vq$ in all Green functions.

As long as the momenta associated with the electron--spin-wave interaction are much smaller than the Fermi momentum, $q \ll k_{\rm F}$, one may expand $\epsilon_{|\vk - \vq|} \approx \epsilon_k - {\hbar \vv_\vk\cdot\vq}$. This approximation allows to perform all sums over electronic momenta (except the interaction momentum $\hbar\vq$) in $\delta\sigma_{\alpha\beta}^{\rm int}$ analytically. Such a calculation was performed in Ref.\ \onlinecite{PhysRevB.64.214204} for a two-dimensional system. Performing the calculation for three dimensions and specializing to the present context gives $\delta \sigma_{\alpha\beta}^{\rm int} = \delta \sigma^{\rm int} \delta_{\alpha\beta}$ with
\begin{align}
\frac{\delta\sigma^{\rm int}}{\sigma_0} = & \frac{2J^2\tau_{\rm el}^2}{\hbar{\cal V}} \sum_\vq \int\frac{d\Omega}{2\pi}
\left[ \frac{\partial}{\partial \Omega}\Omega\coth\left(\frac{\hbar\Omega}{2T}\right)\right] \nonumber\\
& \times {\rm Im}  \Bigg\{
\frac{(\Gamma -1)^2 (\Gamma +1)}{\Gamma }
-\frac{2 \Gamma }{S^2}
\nonumber\\ & \hspace{1.2cm}
+\frac{[1+(1-\Gamma ) i w ]^2}{v^2}
-\frac{\Gamma ^2\left(1+iw \right)}{S^4}
\nonumber\\ & \hspace{1.2cm}
+\frac{\Gamma }{v^2}
\bigg(
1
+\frac{[\Gamma -1][\Gamma -1- i w ]}{\Gamma }
\nonumber\\ & \hspace{1.2cm}
-\frac{\Gamma  [1+i w]}{S^2 }
\bigg)^2\Bigg\}\chi_{-+}^A(\vq,\Omega),\label{eq:res3d}
\end{align}
where we abbreviated
\begin{align}
S & = \sqrt{[1+iw]^2 + v^2}, \qquad
\Gamma = \frac{v}{v- \arctan \frac{v}{1+iw}},
  \label{eq:SGamma}
\end{align}
with
\begin{align}
v & = v_{\rm F}q\tau_{\rm el}, \qquad w = (\Omega+\Delta/\hbar)\tau_{\rm el}. \label{eq:vw}
\end{align}
In Eq.\ (\ref{eq:res3d}) the interaction correction is normalized with respect to the Drude conductivity $\sigma_0 = e^2\nu D$.

The expression still contains a sum over $\vq$ and an integral over $\Omega$, which generally cannot be performed analytically. Therefore, we will focus in the next Section on two different limits in which we can simplify considerably: (i) the clean limit $\Delta\tau_{\rm el}/\hbar \gg 1$, for which we can arrive at an analytic expression for $\delta\sigma^{\rm int}$, and (ii) the dirty limit, where the diffusion approximation still holds.

Note that the result (\ref{eq:res3d}) is still very general: It can be applied to other types of interaction by substituting $J^2\chi^A_{-+}(\vq,\Omega)$ with any advanced interaction propagator of choice, the only requirement being that the interaction is isotropic in $\vq$. Equation (\ref{eq:res3d}) then presents the first-order correction $\delta\sigma^{\rm int}$ resulting from the Fock diagrams shown in Fig.~\ref{fig:dia2}.

\section{Results}\label{sec:res}

\subsection{Clean limit}\label{sec:clean}

The clean limit $\Delta\tau_{\rm el}/\hbar$ is most relevant for the elemental ferromagnetic metals. Indeed, typical electronic elastic scattering rates in disordered metals are of the order $\sim 100$~K (in temperature units), whereas the exchange splitting for most common ferromagnets is typically an order of magnitude larger. In addition to taking the limit $\Delta\tau_{\rm el}/\hbar \gg 1$ we will use two further assumptions to simplify the general result (\ref{eq:res3d}): (i) We assume that the exchange energy is large enough such that $\Delta \gg \hbar q v_{\rm F}$, where $q$ is a typical wave number involved in the interactions. This creates a restriction for the temperature, and limits the results presented in this Section to $T\ll \Delta^2 D^{\rm sw}/\hbar v_{\rm F}^2 \sim \Delta^3/E_{\rm F}^2$, see Ref.\ \onlinecite{1367-2630-15-12-123036} and the discussion below; (ii) We assume that the exchange splitting $\Delta$ is much larger than all relevant spin-wave energies. We will see that this implies $T \ll \Delta$, which is automatically satisfied when $T\ll \Delta^3/E_{\rm F}^2$.

Under these assumptions, we can expand the result (\ref{eq:res3d}) in large $\Delta\tau_{\rm el}/\hbar$. The leading-order contribution is of order $(\hbar/\Delta\tau_{\rm el})^4$ and contains contributions from diagrams 1, 3--8, 11, and 13. The result is
\begin{align}
\frac{\delta\sigma^{\rm int}}{\sigma_0} = \frac{4 J^2 \bar s}{\cal V} & \sum_\vq 
\frac{8 \hbar^2 v_{\rm F}^2 q^2}{9 \Delta^4} \nonumber\\
& \times\left[ \frac{\partial}{\partial \Omega}\Omega\coth\left(\frac{\hbar\Omega}{2T}\right)\bigg|_{\Omega = \omega_\vq^{\rm sw}}-1\right].\label{eq:largd}
\end{align}
The subtraction of the constant $1$ in the summand takes care of the unphysical divergence of $\delta\sigma^{\rm int}$ from large spin-wave frequencies\cite{1367-2630-15-12-123036} and amounts to the change $\delta \sigma \equiv \delta \sigma(T) - \delta \sigma(0)$ of the conductivity correction. The summand of (\ref{eq:largd}) becomes exponentially small for $\hbar\omega_{\bf q}^{\rm sw} \gtrsim T$, which justifies our assumption $\Delta \gg \hbar\Omega$ as long as $T \ll \Delta$. It also introduces $q_{\rm max} = (T/\hbar D^{\rm sw})^{1/2}$ as the scale of the largest wave vector which has to be taken into account. This leads to the restriction $T \ll \Delta^2 D^{\rm sw}/\hbar v_{\rm F}^2$ for the validity of the results presented here, as we anticipated above.\cite{1367-2630-15-12-123036} We also note that our result (\ref{eq:largd}) does not depend on $\tau_{\rm el}$. This indicates that the rapid dephasing of electronic propagators with opposite spin causes the spin-wave-mediated electron--electron interaction to be very short-ranged, shorter than the electronic mean free path $l_{\rm el} = v_{\rm F} \tau_{\rm el}$.

For samples in which all dimensions exceed $q_{\rm max}^{-1}$, the sum over the spin-wave wave vectors ${\bf q}$ in (\ref{eq:largd}) can be converted to a three-dimensional integral. If one or two of the sample dimensions are small, $a \ll q_{\rm max}^{-1}$, the sample can be treated as being effectively two- or one-dimensional, as far as the spin waves are concerned. We see that, in terms of temperature, this regime is reached when $T \ll \hbar D^{\rm sw}/a^2$. For systems with an effective (spin-wave) dimension $d < 3$ one finds effective coupling parameters $J \to J/a^{3-d}$ and $\bar s \to \bar s a^{3-d}$, which have correspondingly changed units of energy times area/length and polarization per area/length for $d=2$, $1$ respectively. We note that, whereas for lower-dimensional samples we assume $a$ to be small, we always take $a$ larger than the electronic mean free path $l_{\rm el}$ so that, from the point of view of electronic impurity scattering, the sample is effectively three-dimensional.\cite{cond_corr_fm:2}

For the spin-wave dispersion of Eq.\ (\ref{eq:disp}) the effectively $d$-dimensional integral over $\vq$ can be evaluated explicitly, and one finds
\begin{align}
 \frac{\delta\sigma^{\rm int}}{\sigma_0} =-& \frac{64 \hbar^2 J^2\bar s v_{\rm F}^2}{9 \Delta^4} \left( \frac{T}{4\pi \hbar D^{\rm sw}} \right)^{\!\!\tfrac{d+2}{2}} \nonumber\\
 &\times \bigg[ 2d\frac{C}{T} {\rm Li}_{\frac{d}{2}}(e^{-\frac{C}{T}}) +d^2{\rm Li}_{\frac{d+2}{2}}(e^{-\frac{C}{T}})\bigg],
 \label{eq:ldli}
\end{align}
where ${\rm Li}_n(z)$ denotes the polylogarithm.
\begin{figure}[t]
\begin{center}
\includegraphics{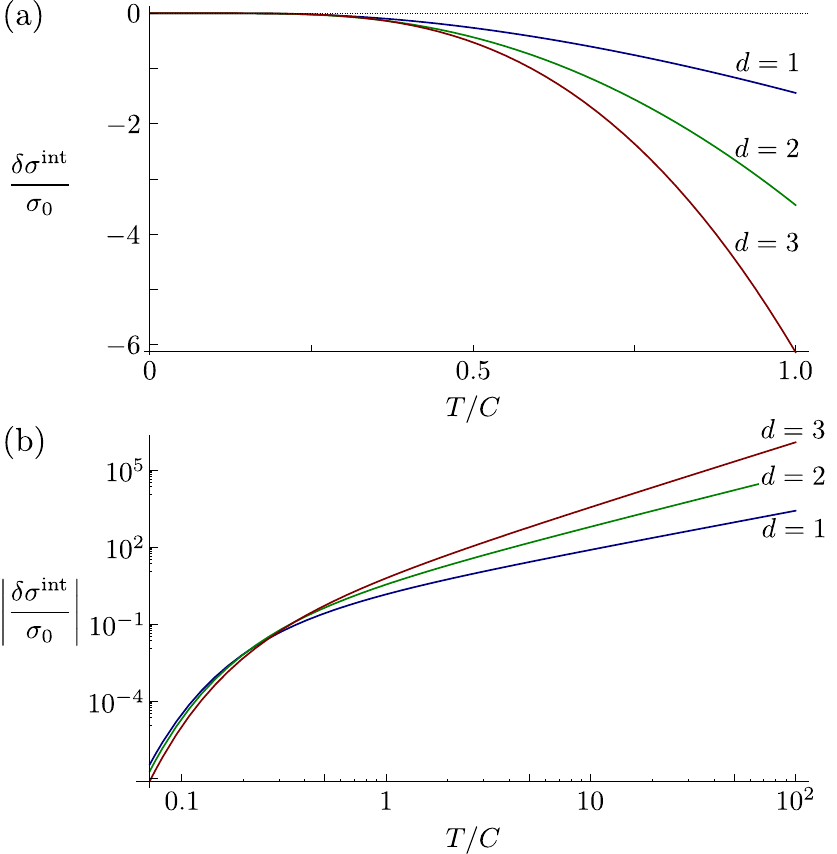}
\caption{(Color online) (a) The spin-wave-induced interaction correction to the conductivity for $d=1,2,3$. The correction is shown in units of $(64 \hbar^2 J^2\bar s v_{\rm F}^2 / 9 \Delta^4)(C / 4\pi \hbar D^{\rm sw})^{(d+2)/2}$. (b) The same three plots, but now on logarithmic scales to show the power-law dependence on $T$ for higher temperatures.}\label{fig:plot1}
\end{center}
\end{figure}
In the limit that the temperature $T$ is much smaller than the spin-wave gap $C$ the result (\ref{eq:ldli}) can be expanded, leading to
\begin{align}
  \frac{\delta\sigma^{\rm int}}{\sigma_0} & = -\frac{128}{9} d \frac{\hbar^2J^2\bar sv_{\rm F}^2}{\Delta^4} \left( \frac{T}{4\pi \hbar D^{\rm sw}} \right)^{\!\!\tfrac{d+2}{2}}\frac{C}{T} e^{-\frac{C}{T}}.
\end{align}
In the opposite regime of large $T/C$ we find
\begin{align}
  \frac{\delta\sigma^{\rm int}}{\sigma_0} & = -\frac{64}{9}d^2 \zeta(\tfrac{d+2}{2}) \frac{\hbar^2J^2\bar sv_{\rm F}^2}{\Delta^4} \left( \frac{T}{4\pi \hbar D^{\rm sw}} \right)^{\!\!\tfrac{d+2}{2}}, \label{eq:cleanres}
\end{align}
where $\zeta(z)$ is the zeta-function. For $d=1,2,3$ the dimension-dependent factor $d^2 \zeta(\tfrac{d+2}{2})$ is $\zeta(\tfrac{3}{2}) \approx 2.61$, $4 \zeta(2) = \tfrac{2}{3}\pi^2 \approx 6.58$, and $9\zeta(\tfrac{5}{2}) \approx 12.1$. We conclude that for temperatures below the spin-wave gap $C$ the interaction correction to the conductivity is suppressed exponentially, and that for large enough temperatures the correction approaches a power-law behavior with power $(d+2)/2$. The full solution (\ref{eq:ldli}) for $d=1,2,3$ is shown in Fig.~\ref{fig:plot1}, which shows $\delta\sigma^{\rm int}/\sigma_0$ in units of $(64 \hbar^2 J^2\bar s v_{\rm F}^2 / 9 \Delta^4)(C / 4\pi \hbar D^{\rm sw})^{(d+2)/2}$ on (a) normal and (b) logarithmic scales. The suppression of the correction at $T\lesssim C$ and the approximate power-law dependence at $T\gtrsim C$ can clearly be recognized in the figures.

\subsection{Dirty limit}

Outside the clean-limit regime $\Delta\tau_{\rm el} /\hbar \gg 1$ we in general cannot obtain closed-form expressions for the conductivity correction $\delta \sigma^{\rm int}$, so that we have to resort to a numerical evaluation of the integrals in Eq.~ (\ref{eq:res3d}). The expressions to evaluate are the simplest in the dirty limit $\Delta\tau_{\rm el}/\hbar \ll 1$, which we discuss here. 

If not only the exchange splitting but also the temperature is much smaller than the electronic elastic scattering rate, $T \ll \hbar/\tau_{\rm el}$, then we may expand Eq.\ (\ref{eq:SGamma}) for small $v$ and $w$. This gives
\begin{align}
\frac{\delta\sigma^{\rm int}}{\sigma_0} = \frac{4J^2\bar s}{\hbar{\cal V}}\int & \frac{d\Omega}{2\pi} \left[ \frac{\partial}{\partial\Omega} \Omega \coth \left(\frac{\hbar\Omega}{2T}\right)-{\rm sign}(\Omega)\right] \nonumber\\
&\times  {\rm Im} \sum_{\bf q} \frac{4}{3} \frac{ Dq^2 {\cal D} ({\bf q},\Omega+\Delta/\hbar)^3}{\hbar\Omega + \hbar\omega_{\bf q}^{\rm sw} + i\eta},
\label{eq:conddif}
\end{align}
where
\begin{equation}
{\cal D}({\bf q},\Omega) = \frac{1}{D q^2 - i\Omega}
\end{equation}
is the diffuson propagator and $D=v_{\rm F}^2\tau_{\rm el}/d$ the diffusion constant. We again subtracted the zero-temperature correction to avoid possible divergences. A closer inspection of the expansion leading to Eq.~(\ref{eq:conddif}) shows that in this case the diagrams 2, 9, and 10 are the dominating ones. Indeed, with $v, w, \Delta\tau_{\rm el}/\hbar \ll 1$ we entered the truly diffusive regime and the largest contributions come as usual from the diagrams with the largest number of diffuson ladders.

\begin{figure}[t]
\begin{center}
\includegraphics{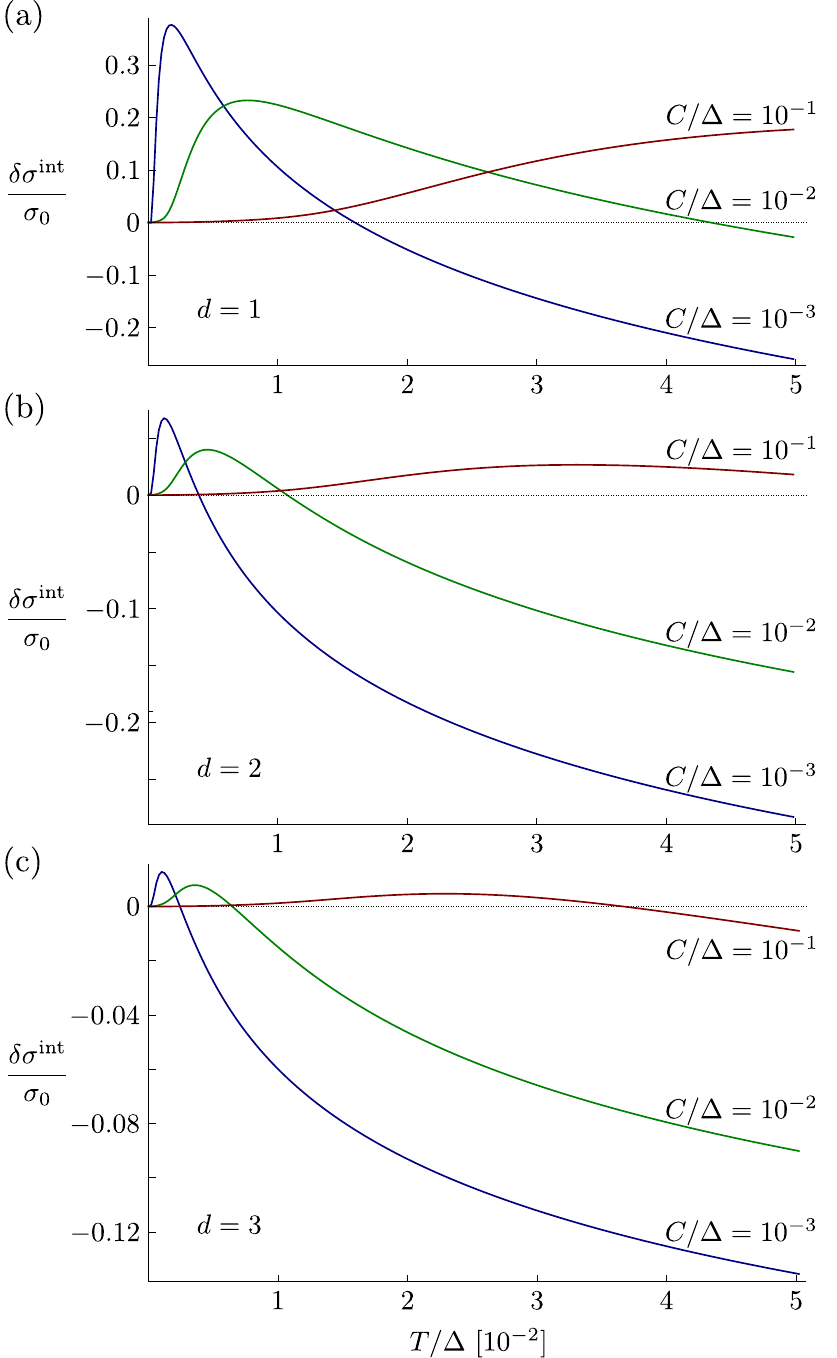}
\caption{(Color online) The spin-wave-mediated interaction correction to the conductivity for a ferromagnet in the dirty limit. The correction for effectively (a) one-dimensional, (b) two-dimensional, and (c) three-dimensional systems is shown in units of $(J^2\bar s/\Delta^2)(\Delta/\hbar D^{\rm sw})^{d/2}$ for three different ratios of the spin-wave gap $C$ and the exchange splitting $\Delta$. For all curves we have set $D^{\rm sw}/D = 10^{-3}$.}\label{fig:plot2}
\end{center}
\end{figure}
If one, two, or three of the dimensions of the sample are large enough, i.e., much larger than $q_{\rm max}^{-1}$, we can rewrite the sum over $\vq$ as an integral in the same way as we did in the previous Section.\cite{cond_corr_fm:3} The resulting integrals can be evaluated numerically, and in Fig.~\ref{fig:plot2} we show the resulting $\delta \sigma^{\rm int}$ for an effectively (a) one-dimensional, (b) two-dimensional, and (c) three-dimensional sample. For all curves we have set $D^{\rm sw}/D = 10^{-3}$, and we show results for three different ratios $C/\Delta$. We note that in this diffusive limit the correction $\delta\sigma^{\rm int}$ shows qualitatively the same behavior as the spin-wave-induced correction to the electronic density of states.\cite{1367-2630-15-12-123036}

\section{Conclusion}\label{sec:conc}

In this article, we calculated the interaction quantum correction---the ``Altshuler-Aronov correction''---to the conductivity of a disordered ferromagnetic metal that is due to spin-wave-mediated electron--electron interactions. We used a simple model in which electrical current is carried by $s$-band conduction electrons and spin waves exist as fluctuations of the magnetization of $d$-band electron spins. The exchange interaction between the $s$- and $d$-band electrons allows for dynamical processes in which the conduction electrons excite or absorb spin waves, leading to an effective spin-wave-mediated interaction.

Our main result, the most general result obtainable within diagrammatic perturbation theory {\em without} making assumptions on the relative magnitudes of the exchange splitting $\Delta$ and the elastic mean-free time $\tau_{\rm el}$, is given in Eq.\ (\ref{eq:res3d}). The only assumption for this result is that $\Delta \ll E_{\rm F}$, which is a necessary requirement for the applicability of diagrammatic perturbation theory. We then simplified this result for the limiting cases of a ``clean'' ferromagnet ($\Delta\tau_{\rm el}/\hbar \gg 1$) and a ``dirty'' ferromagnet ($\Delta\tau_{\rm el}/\hbar \ll 1$). For the clean case, which is the most relevant for the elemental ferromagnets, the correction $\delta\sigma^{\rm int}$ is suppressed exponentially for temperatures below the spin-wave gap, whereas at higher temperatures it acquires a temperature dependence of $\propto T^{d/2+1}$, where $d$ is the effective dimensionality of the spin-waves. The results found in this limit do not depend on $\tau_{\rm el}$, indicating that the effective interactions are short-range (shorter than the mean free path $l_{\rm el}$). For the dirty limit, the correction has a non-monotonous temperature dependence, qualitatively resembling the spin-wave-induced correction to the density of states.\cite{1367-2630-15-12-123036}

A relevant question is how our results compare to similar calculations existing in the literature. To our knowledge, the only theoretical work addressing the same quantum correction is an article by Misra {\em et al.}, the second part of which addresses the spin-wave-mediated quantum correction to the conductivity.\cite{PhysRevB.79.140408} There, the authors find for $d=2$ a positive, linear-in-$T$ correction, which they state to be valid in the clean limit $\Delta\tau_{\rm el}/\hbar \gg 1$. These results clearly do not agree with the results of our Sec.~\ref{sec:clean}, which has a negative correction proportional to $T^2$. We attribute the disagreement to the (implicit) use of a standard diffusion approximation in Ref.\ \onlinecite{PhysRevB.79.140408}, which leads the authors to keep only those diagrams from Fig.~\ref{fig:dia3}(b) with the largest number of diffusons (diagrams 2, 9, and 10), since those are assumed to be the ones that explore the largest phase space and therefore yield the largest contributions to $\delta\sigma$. At that point the clean-limit assumption $\Delta\tau_{\rm el}/\hbar \gg 1$ is made and the results are simplified accordingly, leading to a correction linear in $T$ with a positive sign. However, as we argued above, in the limit of large $\Delta\tau_{\rm el}/\hbar$, the diffuson ladders in the diagrams dephase on a length scale so short that they do not increase the phase space explored as compared to the diagrams without diffuson ladders. We thus believe that the correct approach in this limit is to keep all 14 diagrams of Fig.~\ref{fig:dia3}(b) and expand them consistently in orders of $\hbar/\Delta\tau_{\rm el}$.

\begin{figure}[t]
\begin{center}
\includegraphics{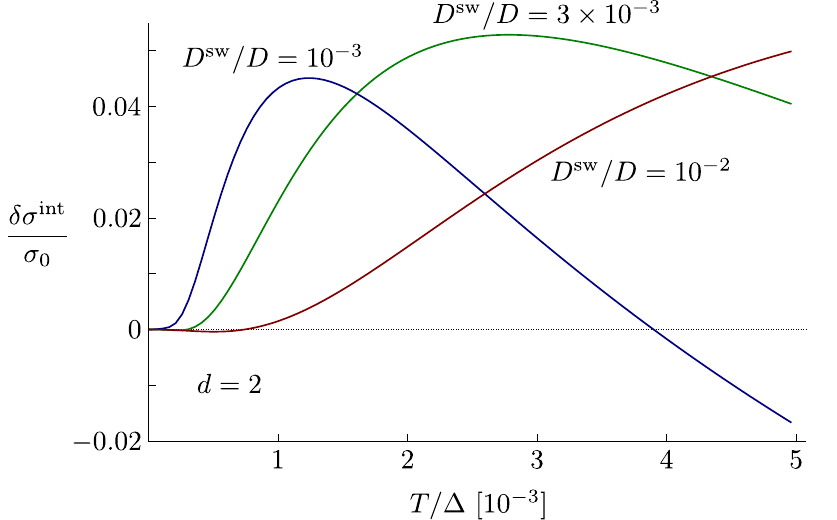}
\caption{(Color online) The interaction correction to the conductivity in the diffusive regime at very low temperatures. The correction is shown in units of $(J^2\bar s/\Delta^2)(\Delta/\hbar D^{\rm sw})^{d/2}$, now for three different ratios $D^{\rm sw}/D$. For all curves we have set $C/\Delta = 10^{-3}$.}\label{fig:plot3}
\end{center}
\end{figure}
Our numerical results indicate that a {\it positive} correction $\delta \sigma^{\rm int}$ only appears in the dirty regime at the low-temperature side. It would be interesting to understand to what extent this regime overlaps with the regime investigated in Ref.~\onlinecite{PhysRevB.79.140408}. In Fig.~\ref{fig:plot3} we thus show $\delta\sigma^{\rm int}$ for temperatures a factor 10 smaller than in Fig.~\ref{fig:plot2}. We focus on $d=2$, fix $C/\Delta = 10^{-3}$, and show $\delta\sigma^{\rm int}$ for three different ratios of $D^{\rm sw}/D$. We see that there is a regime where $\delta\sigma^{\rm int}$ is positive and increases with temperature, roughly when $T/\Delta \lesssim D^{\rm sw}/D$. All thermal spin waves have $\hbar D^{\rm sw} q^2 \lesssim T$ (assuming a vanishingly small spin-wave gap), and if $T/\Delta \lesssim D^{\rm sw}/D$ then this causes the diffuson modes excited by thermal spin waves to have $\hbar D q^2 \lesssim \Delta$. A simplification of (\ref{eq:conddif}) with $\Delta\tau_{\rm el}/\hbar \gg 1$ would indeed be consistent with neglecting $\hbar D q^2$ compared to $\Delta$. From our results however, we have no indication that the correction has a linear $T$-dependence over an extended range of temperatures.

We note that the linear contribution to $\sigma (T)$ observed in the experiments of Ref.~\onlinecite{PhysRevB.79.140408} could possibly be explained as a precursor of a transition into the ballistic regime. For ballistic transport, with $T\tau_{\rm el}/\hbar \gg 1$, the (regular) Coulomb Altshuler-Aronov correction is known to become linear in temperature for effectively two-dimensional systems. The sign of this contribution was found to depend on the strength of the interaction:\cite{PhysRevB.64.214204} For weak interaction one finds a linear contribution to $\sigma(T)$ with a positive sign, as observed experimentally in Ref.~\onlinecite{PhysRevB.79.140408}.

We finally compare the order of magnitude of the spin-wave-induced Altshuler-Aronov correction we found to that of the Coulomb interaction correction. We set $d=3$, for which the latter reads in the diffusive limit\cite{aa:book}
\begin{align}
\frac{\delta\sigma^{\rm int,ee}}{\sigma_0} &= \frac{A}{2\pi^2 \nu \hbar D}\sqrt{\frac{T}{\hbar D}} \nonumber \\ &\sim \sqrt{\frac{T}{E_{\rm F}}} \left(\frac{\hbar}{E_{\rm F}\tau_{\rm el}}\right)^{3/2},
\end{align}
where $A \approx 0.61$ is a numerical factor. We compare this correction to our clean result of Sec.~\ref{sec:clean}. For $d=3$ we see that the spin-wave-induced correction tends to be $\propto T^{5/2}$, which indicates that this correction could become dominant for higher temperatures. We thus assume $T\gg C$, for which we find
\begin{align}
\frac{\delta\sigma^{\rm int,sw}}{\sigma_0} \sim \left(\frac{T}{\Delta}\right)^{5/2}\left(\frac{E_{\rm F}}{\Delta}\right)^2,
\end{align}
where we assumed that $J\bar s \sim \Delta$, $D^{\rm sw} \sim \Delta/\hbar k_{\rm F}^2$, and $\bar s \sim k_{\rm F}^3$. A straightforward comparison of the two corrections yields a minimum temperature for the spin-wave-induced corrections to dominate,
\begin{align}
T_{\rm min} \sim \left(\frac{\hbar}{\Delta\tau_{\rm el}}\right)^{3/4} \frac{\Delta^3}{E_{\rm F}^2}.\label{eq:tmin}
\end{align}
Taking parameters from Refs.~\onlinecite{ashmer} and \onlinecite{PhysRevB.66.024433} for iron ($\Delta \approx 9000$~K, $E_{\rm F} \approx 11$~eV) we find for $\tau_{\rm el} = 10^{-14}$~s$^{-1}$ a minimum temperature of $T_{\rm min} \approx 7$~K. This in principle allows for a significant regime of temperatures where the spin-wave-induced Altshuler-Aronov correction could be the dominant quantum correction to the conductivity. We caution that the results from Sec.~\ref{sec:clean} were derived under the assumption that $T$ is much smaller than $\Delta^3/E_{\rm F}^2$, which is only a factor $(\Delta\tau_{\rm el}/\hbar)^{1/4}$ larger than $T_{\rm min}$. This means that the estimate (\ref{eq:tmin}) has to be interpreted as of a rather qualitative nature.

\acknowledgments

We acknowledge very helpful discussions with Georg Schwiete, Martin Schneider, and Karsten Flensberg. This work was supported by the Alexander von Humboldt Foundation in the framework of the Alexander von Humboldt Professorship program, endowed by the Federal Ministry of Education and Research.


\begin{thebibliography}{10}
\bibitem{cond_corr_fm:1}
The exchange splitting
  $\Delta$ can differ from $J \bar{s}$ if exchange interactions between
  conduction electrons are taken into account. We will thus not make use of the
  relation $\Delta = 2J \bar{s}$, but keep $\Delta$ as an independent parameter
  instead.
\bibitem{note_diagrams}
At first sight only diagrams 2, 9, and 10 seem to be relevant, since those are the only ones containing three diffuson ladders. These diagrams however contain scattering events that, to first approximation, randomize the relative direction of the momenta at the two current vertices. To get a finite contribution from these diagrams, one needs to track the correlation between the two momenta by expanding the short-range Green functions in small $\vq$. Since this yields a small factor, one should also include the leading diagrams without scattering events connecting the upper and lower Green functions: diagrams 1, 5--7 (describing a Hikami box), and 8.
\bibitem{cond_corr_fm:2}
For a strictly two-dimensional sample, one should use (3.26) from
  Ref.~\onlinecite{PhysRevB.64.214204}. Expansion in $\hbar/\Delta\tau_{\rm
  el}$ then yields exactly the same expression as (\ref{eq:largd}) but with the
  numerical factor $\tfrac{8}{9}$ replaced by $\tfrac{3}{2}$: The result is
  larger by a factor $\tfrac{27}{16}$.
\bibitem{cond_corr_fm:3}
For a truly two-dimensional sample, i.e., if one dimension is smaller than
  $l_{\rm el}$, one can still use Eq.~(\ref{eq:conddif}) but one has to replace
  the factor $4/3$ by $2$.
\end{thebibliography}

\end{document}